# Remarks on the Criteria of Constructing MIMO-MAC DMT Optimal Codes

Hsiao-feng (Francis) Lu, Jyrki Lahtonen, Roope Vehkalahti, Camilla Hollanti

*Abstract*— In this paper we investigate the criteria proposed by Coronel *et al.* for constructing MIMO MAC-DMT optimal codes over several classes of fading channels. We first give a counterexample showing their DMT result is not correct when the channel is frequency-selective. For the case of symmetric MIMO-MAC flat fading channels, their DMT result reduces to exactly the same as that derived by Tse *et al.* , and we therefore focus on their criteria for constructing MAC-DMT optimal codes, especially when the number of receive antennas is sufficiently large. In such case, we show their criterion is equivalent to requiring the codes of any subset of users to satisfy a joint non-vanishing determinant criterion when the system operates in the antenna pooling regime. Finally an upper bound on the product of minimum eigenvalues of the difference matrices is provided, and is used to show any MIMO-MAC codes satisfying their criterion can possibly exist only when the target multiplexing gain is small.

## I. Introduction

Consider a MIMO multiple-access (MAC) channel with $U$ users, each equipped with $n_t$ transmit antennas. The receiver at the base-station is assumed to have $n_r$ receive antennas. Code matrices of each user are transmitted synchronously and independently in $M$ channel uses. Let $H_{u,m}$ denote the channel matrix of the $u$th user at the $m$th channel use; then given transmitted signal $\underline{x}_{u,m}$, the received signal vector is

$$\underline{y}_m = \sqrt{\frac{\text{SNR}}{n_t}} \sum_{u=1}^{U} H_{u,m}\, \underline{x}_{u,m} + \underline{z}_m, \quad m=0,1,\cdots,M-1, \tag{1}$$

where $\underline{z}_m$ is the noise vector with i.i.d. $\mathbb{CN}(0,1)$ entries at the $m$th channel use. The transmitted signal $\underline{x}_{u,m}$ is required to satisfy the power constraint of $\mathbb{E}\left\Vert \underline{x}_{u,m}\right\Vert^2 = 1$.

Following [1], the MIMO-MAC channel is assumed to be frequency selective and has the following assumptions:

1) the entries of the channel matrix $H_{u,m}$ are i.i.d. $\mathbb{CN}(0,1)$ complex Gaussian random variables,
2) the channels correspond to different users are statistically independent, and
3) the channels $\{H_{u,m}: m=0,\cdots,M-1\}$ seen by the $u$th user can be correlated in time.

Due to possible time-correlation, Coronel *et al.* [1] defined an $(M \times M)$ matrix $R_H$ whose entries are given by

$$\mathbb{E}\left\{H_{u,m}(i,j) H_{u',m'}(i,j)\right\} = R_H(m,m')\delta_{u,u'}$$

to model the time correlations, where by $H_{u,m}(i,j)$ we mean the $(i,j)$th entry of the channel matrix $H_{u,m}$.

Let $\mathcal{X}_u = \{X_u\}$ be the block length $M$ code of user $u$, consisting of $(n_t \times M)$ code matrices satisfying the average power constraint $\mathbb{E}\left\Vert X_u\right\Vert_F^2 \leq n_t M$. The codebook $\mathcal{X}_u$ has size $|\mathcal{X}_u| \doteq \text{SNR}^{Mr_u}$ such that the user $u$ transmits at multiplexing gain $r_u$.

Let $\mathcal{U} = \{1,2,\cdots,U\}$ denote the set of all users. For any subset of users, $\mathcal{S} \subseteq \mathcal{U}$, the probability of users in $\mathcal{S}$ being in outage is lower bounded by

$$\Pr\{\mathcal{O}_{\mathcal{S}}\} \dotgeq \text{SNR}^{-d_{\mathcal{S}}(r(\mathcal{S}))}$$

where $r(\mathcal{S}) := \sum_{u \in \mathcal{S}} r_u$. For integral values of $r(\mathcal{S})$, Coronel *et al.* showed [1] that

$$d_{\mathcal{S}}(r(\mathcal{S})) := [m(\mathcal{S}) - r(\mathcal{S})][\rho M(\mathcal{S}) - r(\mathcal{S})]$$

where $\rho = \text{rank}(R_H)$, $m(\mathcal{S}) := \min\{|\mathcal{S}|n_t, n_r\}$, and $M(\mathcal{S}) := \max\{|\mathcal{S}|n_t, n_r\}$, and that $d_{\mathcal{S}}(r(\mathcal{S}))$ is a piecewise linear function for non-integral values of $r(\mathcal{S})$

For any specific choice of codebook $\mathcal{X}_u$, Coronel *et al.* [1] studied the error performance of such code and provided a criterion based on the eigenvalues of the difference code matrices such that $\mathcal{X}_u$ has error probability upper bounded by the outage probability. We reproduce their result in the theorem below.

*Theorem 1 ( [1], [2]):* For every $u \in \mathcal{S} \subseteq \mathcal{U}$, let $\mathcal{X}_u$ have block length $M \geq \rho|\mathcal{S}|n_t$. Let the nonzero eigenvalues of $R_H \odot \left(\sum_{u \in \mathcal{S}} E_u^{\dagger} E_u\right)$, where $E_u = X_u - X'_u$ and $X_u \neq X'_u \in \mathcal{X}_u$ be given in ascending order - at every SNR level by $\lambda_k$, $k=1,2,\cdots,\rho|\mathcal{S}|n_t$. Furthermore, set

$$\Lambda_{m(\mathcal{S})}^{\rho|\mathcal{S}|n_t}(\text{SNR}) := \min_{E_u:u\in\mathcal{S}} \prod_{k=1}^{m(\mathcal{S})} \lambda_k. \tag{2}$$

If there exists an $\epsilon > 0$ independent of SNR and $r(\mathcal{S})$ such that

$$\Lambda_{m(\mathcal{S})}^{\rho|\mathcal{S}|n_t}(\text{SNR}) \dotgeq \text{SNR}^{-(r(\mathcal{S})-\epsilon)}, \tag{3}$$

then under ML decoding, the event $\mathcal{E}_{\mathcal{S}}$ of the users of $\mathcal{S}$ in error has probability upper bounded by $P(\mathcal{E}_{\mathcal{S}}) \dotleq \text{SNR}^{-d_{\mathcal{S}}(r(\mathcal{S}))}$. ∎

They then derived the optimal MAC-DMT tradeoff of this MIMO-MAC frequency selective fading channel and at the same time, provided a sufficient criterion for codes to achieve such MAC-DMT.

*Theorem 2 ( [1]):* The optimal tradeoff of the frequency selective-fading MIMO MAC channel defined in (1) is given by $d^*(\underline{r}) = d_{\mathcal{S}^*}(r(\mathcal{S}^*))$, where

$$\mathcal{S}^* := \arg\min_{\mathcal{S} \subseteq \mathcal{U}} d_{\mathcal{S}}(r(\mathcal{S}))$$

is the dominant outage set for error performance. That is, the optimal MAC-DMT is given by

$$d^*(\underline{r}) = (m(\mathcal{S}^*) - r(\mathcal{S}^*))(\rho M(\mathcal{S}^*) - r(\mathcal{S}^*)), \tag{4}$$

where $\underline{r} = (r_1, \cdots, r_U)$ is the vector of the multiplexing gains of all users. Moreover, if the overall family of codes $\mathcal{X} = \mathcal{X}_1 \times \cdots \times \mathcal{X}_U$ satisfies (3) for the dominant outage set $\mathcal{S}^*$ and for every $\mathcal{S} \neq \mathcal{S}^*$, there exists $\epsilon > 0$ such that

$$\Lambda_{m(\mathcal{S})}^{\rho|\mathcal{S}|n_t} (\text{SNR}) \overset{.}{\geq} \text{SNR}^{-(\gamma_\mathcal{S} - \epsilon)} \quad (5)$$

where

$$0 \leq \gamma_\mathcal{S} \leq \Gamma_\mathcal{S} := d_\mathcal{S}^{-1}(d_{\mathcal{S}^*}(r(\mathcal{S}^*))) \quad (6)$$

then $\mathcal{X}$ achieves the optimal DMT $d^*(\underline{r})$. ∎

We remark that in a preceding publication [2] Coronel *et al.* had provided a different criterion that is stronger than condition (5). The difference between this criterion and (5) is that the former replaces $\gamma_\mathcal{S}$ by $r(\mathcal{S})$. Obviously since $\mathcal{S}^*$ is the dominant outage set, we have $r(\mathcal{S}) \leq \Gamma_\mathcal{S}$ for all $\mathcal{S} \subseteq \mathcal{U}$, and (5) is more relaxed than that in [2]. Moreover, we note that (5) can be further relaxed by replacing $\gamma_\mathcal{S}$ with $\Gamma_\mathcal{S}$. As a result, in all subsequent discussions we will use $\Gamma_\mathcal{S}$ instead of $\gamma_\mathcal{S}$.

*A. DMT of Frequency Selective Channels: A Counterexample*

Unfortunately, the MAC-DMT (4) claimed by Coronel *et al.* [1] for the frequency selective fading channel is not correct. This can be seen from the counterexample below.

*Example 1:* Consider a single-user, point-to-point, multi-block fading MIMO channel [3], [4]. In terms of the channel model (1) we assume the channel has the following input-output relation

$$\underline{y}_m = \sqrt{\frac{\text{SNR}}{n_t}} H_m \underline{x}_m + \underline{z}_m, \quad m = 0, 1, \cdots, M-1$$

where $M = 2n_t$ and the channel matrices are given by

$$H_m = \begin{cases} H_0, & m = 0, 1, \cdots, n_t - 1 \\ H_{n_t}, & m = n_t, \cdots, 2n_t - 1 \end{cases}. \quad (7)$$

The entries of $H_0$ and $H_{n_t}$ are i.i.d. $\mathbb{CN}(0,1)$ random variables. Thus (7) models a quasi-static MIMO Rayleigh block fading channel in which coding is allowed to be spread over two independent fading blocks, each fixed for $n_t$ channel uses. Given multiplexing gain $r$, it is well known [3], [4] that the optimal MAC-DMT of this channel equals

$$d^*(r) = 2(n_t - r)(n_r - r) \quad (8)$$

since the two fading blocks are statistically independent.

On the other hand, from Theorem 2 we note the followings.

1) The time correlation matrix $R_H$ is

$$R_H = \begin{bmatrix} \mathbf{1} & \\ & \mathbf{1} \end{bmatrix}$$

where $\mathbf{1}$ is the all-one matrix of size $(n_t \times n_t)$. Therefore $\rho = \text{rank}(R_H) = 2$.

2) The dominant outage set $\mathcal{S}^* = \{1\}$ since this is a single-user, point-to-point channel. Hence $m(\mathcal{S}^*) = \min\{n_t, n_r\}$ and $M(\mathcal{S}^*) = \max\{n_t, n_r\}$.

Substituting the above into (4) of Theorem 2, Coronel *et al.* claimed however the DMT of this channel is

$$d_{Coronel}^*(r) = (\min\{n_t, n_r\} - r)(2\max\{n_t, n_r\} - r). \quad (9)$$

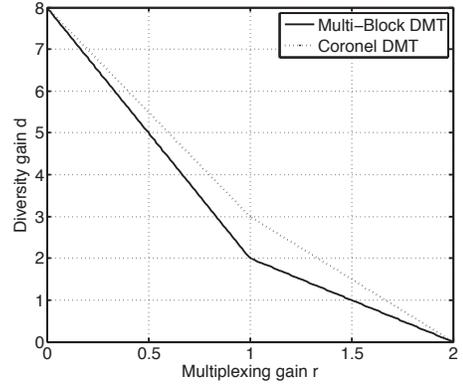

Fig. 1. The two MAC-DMTs (8) (in solid line) and (9) (in dotted line) for single-user multi-block fading channel.

The above disagrees completely with the well-known multi-block DMT result (8) of [3], [4]. In Fig. 1 we plot the two MAC-DMTs, (8) and (9), for the case of $n_t = n_r = 2$. It is seen that the latter (9) is too optimistic on the diversity gain at all values of $r$. Hence we conclude that (4) claimed by Coronel *et al.* is not correct for the case of frequency selective fading channels. ∎

The above example shows the MAC-DMT result (4) claimed by Coronel *et al.* [1], [2] is not correct for the case of single-user MIMO frequency-selective channel. As their result fails in the point-to-point scenario, it will not hold in the MAC case either.

*B. DMT of Flat-Fading MIMO-MAC Channels*

For the flat fading case let us consider a MIMO-MAC channel with $U$ users, communicating independently and synchronously over a quasi-static Rayleigh flat fading channel. Because of the quasi-static assumption we shall have

$$H_{u,m} = H_u, \quad m = 0, 1, \cdots, M-1.$$

Entries of $H_u$ are i.i.d. $\mathbb{CN}(0,1)$ random variables. The time-correlation matrix $R_H$ in this case is given by $R_H = \mathbf{1}_{M \times M}$, where $\mathbf{1}_{M \times M}$ is the all-one matrix of size $(M \times M)$, hence $\rho = \text{rank}(R_H) = 1$.

Here we are interested in the symmetric MIMO-MAC flat fading channel [5] in which all users transmit at the same level of multiplexing gain, i.e. $r_1 = \cdots = r_U = r$. Applying these assumptions to Theorem 2 the resulting MAC-DMT is given by

$$d^*(r) = \min_{1 \leq s \leq U} (sn_t - sr)(n_r - sr)$$

which is exactly the same as that shown by Tse *et al.* [5].

In the remaining of this paper we will focus on the investigation of criterion (5) in Theorem 2 for constructing MAC-DMT optimal codes. Specifically, when $n_r \geq Un_t$,

1) if the system operates in antenna pooling regime, in Section II we will show the relaxation of $\gamma_\mathcal{S}$ in (5) is not possible, and we must have $\gamma_\mathcal{S} = |\mathcal{S}| r$ in (5). Also, we will relate criterion (5) to the non-vanishing determinant (NVD) criterion [6]–[8] that is well-known

for constructing approximately universal point-to-point space-time codes.
2) in Section III, based on an analysis of the minimum determinant we will provide a stronger result that can be applied to all range of multiplexing gain $r$.

## II. CONNECTION TO NVD CRITERION

In this section, we will focus on the case of MIMO-MAC flat fading channels and will assume
1) $n_r \geq Un_t$ and
2) the multiplexing gain $r_u = r$ for all users.

Following Theorem 1, let $\mathcal{X}_u$ be the codebook of the $u$th user with block length $M \geq Un_t$ and consist of $(n_t \times M)$ code matrices. As $R_H = \mathbf{1}_{M \times M}$, for any subset $\mathcal{S}$ of users the matrix in Theorem 1 simplifies to $R_H \odot \left( \sum_{u \in \mathcal{S}} E_u^\dagger E_u \right) = \left( \sum_{u \in \mathcal{S}} E_u^\dagger E_u \right)$, where $E_u = X_u - X_u'$ and $X_u, X_u' \in \mathcal{X}_u$. For the ease of presentation, below we define a notation for concatenating matrices that will be used frequently in the later discussions.

*Definition 1:* Let $X_1, \cdots, X_s$ be matrices with the same number of columns; then we define the vertical concatenation of these matrices as

$$M(X_1, \cdots, X_s) := \begin{bmatrix} X_1 \\ \vdots \\ X_s \end{bmatrix}.$$

With the above notation, set

$$\Delta_\mathcal{S} := M(E_{u_1}, \cdots, E_{u_S}) \quad (10)$$

where $\mathcal{S} = \{u_1, \cdots, u_S\}$; then the nonzero eigenvalues of $\left( \sum_{u \in \mathcal{S}} E_u^\dagger E_u \right)$ are the same as those of $\Delta_\mathcal{S} \Delta_\mathcal{S}^\dagger$. Clearly, $\text{rank}(\Delta_\mathcal{S}) \leq |\mathcal{S}| n_t$. Since $n_r \geq Un_t$ by assumption, we have $m(\mathcal{S}) = |\mathcal{S}| n_t$. As $\Lambda_{m(\mathcal{S})}^{|\mathcal{S}|n_t}(\text{SNR})$ is comprised of the product of the least $m(\mathcal{S})$ nonzero eigenvalues of $\left( \sum_{u \in \mathcal{S}} E_u^\dagger E_u \right)$, (2) forces $\text{rank}(\Delta_\mathcal{S}) = m(\mathcal{S}) = |\mathcal{S}| n_t$. Hence we can rewrite (2) as

$$\Lambda_{m(\mathcal{S})}^{|\mathcal{S}|n_t}(\text{SNR}) = \min_{E_u: u \in \mathcal{S}} \prod_{k=1}^{m(\mathcal{S})} \lambda_k = \min_{E_u: u \in \mathcal{S}} \det\left( \Delta_\mathcal{S} \Delta_\mathcal{S}^\dagger \right),$$

where $\lambda_k$ are the nonzero eigenvalues of $\left( \sum_{u \in \mathcal{S}} E_u^\dagger E_u \right)$, or equivalently, the eigenvalues of $\Delta_\mathcal{S} \Delta_\mathcal{S}^\dagger$, arranged in the ascending order. Moreover, condition (3) can be reformulated as

$$\Lambda_{m(\mathcal{S})}^{|\mathcal{S}|n_t}(\text{SNR}) = \min_{E_u: u \in \mathcal{S}} \det\left( \Delta_\mathcal{S} \Delta_\mathcal{S}^\dagger \right) \,\dot{\geq}\, \text{SNR}^{-(|\mathcal{S}|r - \epsilon)}, \quad (11)$$

and similarly condition (5) can be rewritten as

$$\Lambda_{m(\mathcal{S})}^{|\mathcal{S}|n_t}(\text{SNR}) = \min_{E_u: u \in \mathcal{S}} \det\left( \Delta_\mathcal{S} \Delta_\mathcal{S}^\dagger \right) \,\dot{\geq}\, \text{SNR}^{-(\Gamma_\mathcal{S} - \epsilon)}. \quad (12)$$

Our first goal in this section is to relate the above conditions, (11) and (12), to the well-known NVD condition [6]–[8], [8] for constructing approximately universal space-time codes. To this end, we recall the transmit-receive channel model of [6] is

$$Y = \kappa \sum_{u=1}^{U} H_u C_u + Z \quad (13)$$

where $Y$ is the $(n_r \times M)$ received signal matrix, $H_u$ is the $(n_r \times n_t)$ channel matrix of user $u$, and $C_u \in \mathcal{C}_u$ is the corresponding code matrix with power constraint $\mathbb{E} \|C_u\|^2 \,\dot{\leq}\, \text{SNR}^{\frac{r}{n_t}}$. $\kappa$ is a scaling parameter with $\kappa^2 \,\dot{=}\, \text{SNR}^{1 - \frac{r}{n_t}}$ such that $\mathbb{E} \|\kappa C_u\|^2 \,\dot{=}\, \mathbb{E} \left\| \sqrt{\frac{\text{SNR}}{n_t}} X_u \right\|^2 \,\dot{=}\, \text{SNR}$ and the two models, (1) and (13), agree on the same level of input-SNR.

In [6], [8], the NVD condition states for any single code $\mathcal{C}_u$, if $\min_{D_u} \det\left( D_u D_u^\dagger \right) \geq \text{SNR}^0$ for any $D_u = C_u - C_u'$ and $C_u \neq C_u' \in \mathcal{C}_u$, then the error probability of $\mathcal{C}_u$ is upper bounded by the corresponding outage probability. We remark that this NVD condition can be relaxed such that only exponential inequality $\min_{D_u} \det\left( D_u D_u^\dagger \right) \,\dot{\geq}\, \text{SNR}^0$ is needed without affecting the proof in [6]. The NVD condition can be extended to the MAC case as well (see (15) below).

Contrast to the channel model (1) we see

$$X_u = \text{SNR}^{-\frac{r}{2n_t}} C_u \quad (14)$$

With the above in mind, substituting (14) into (10) gives

$$\Delta_\mathcal{S} = \text{SNR}^{-\frac{r}{2n_t}} \Delta_C$$

where $\Delta_C = M(D_{u_1}, \cdots, D_{u_S})$, $D_u = C_u - C_u'$, and $C_u$ and $C_u'$ are the code matrices associated with $X_u$ and $X_u'$, respectively. Hence

$$\det\left( \Delta_\mathcal{S} \Delta_\mathcal{S}^\dagger \right) = \left[ \text{SNR}^{-\frac{r}{2n_t}} \right]^{2|\mathcal{S}|n_t} \det\left( \Delta_C \Delta_C^\dagger \right).$$

After clearing the common terms, condition (11) is equivalent to

$$\mathfrak{D}(\mathcal{S}) := \min_{D_u: u \in \mathcal{S}} \det\left( \Delta_C \Delta_C^\dagger \right) \,\dot{\geq}\, \text{SNR}^\epsilon \quad (15)$$

for some $\epsilon > 0$.

*Remark 2.1:* We remark that condition (15) is exactly the NVD condition shown in [6], [8]. Hence, Theorem 1 when restricted to the case of flat fading channels, is equivalent to an earlier result of [6], see Theorems 2 and 3 of [6]. ∎

Next, we turn our attention to the condition (12). Again applying the key relation (14) to (12) and after clearing the common terms, condition (12) can be reformulated in terms of the difference matrix $\Delta_C$. Thus in order to achieve the MAC-DMT optimality Theorem 2 requires for every $\mathcal{S} \subseteq \mathcal{U}$

$$\mathfrak{D}(\mathcal{S}) = \min_{D_u: u \in \mathcal{S}} \det\left( \Delta_C \Delta_C^\dagger \right) \,\dot{\geq}\, \text{SNR}^{-(\Gamma_\mathcal{S} - |\mathcal{S}|r - \epsilon)}. \quad (16)$$

For the ease of handling the parameter $\Gamma_\mathcal{S}$, below we will restrict ourselves to the case of $n_r = Un_t$. Recall the symmetric MIMO-MAC DMT of this case [5] is given by

$$d^*(r) = d_{\mathcal{S}^*}\left( r\left( \mathcal{S}^* \right) \right) = \begin{cases} d_1(r), & r \in \left[ 0, \frac{Un_t}{U+1} \right] \\ d_U(Ur), & r \in \left[ \frac{Un_t}{U+1}, n_t \right] \end{cases} \quad (17)$$

where $d_1(r) = (n_t - r)(n_r - r)$ represents the DMT for $\mathcal{S}^* = \{1\}$ when $r \in \left[ 0, \frac{Un_t}{U+1} \right]$. This interval of $r$ is coined the *single-user performance regime* by Tse *et al.* [5]. $d_U(r) = (Un_t - Ur)(n_r - Ur)$ is the DMT for $\mathcal{S}^* = \mathcal{U}$ and dominates the MAC-DMT when $r \in \left[ \frac{Un_t}{U+1}, n_t \right]$. Such regime is called the *antenna-pooling regime*. Here we focus on the latter regime, i.e. the case when $\mathcal{S}^* = \mathcal{U}$, and distinguish two kinds of outage sets.

1) For $\mathcal{S}$ with $|\mathcal{S}| = 1$, it is easy to verify that

$$\Gamma_1 := \Gamma_\mathcal{S}\big|_{|\mathcal{S}|=1} = d_1^{-1}\left(d_U(Ur)\right) > r$$

for $U > 1$ since $r$ in the antenna pooling regime and $d_1(r) > d_U(Ur)$. Hence (16) requires the individual codes $\mathcal{C}_u$ of every user $u$ to satisfy

$$\mathfrak{D}(\{u\}) = \min_{D_u} \det\left(D_u D_u^\dagger\right) \;\dot\geq\; \mathrm{SNR}^{-(\Gamma_1 - r)},$$

i.e. allowing vanishing determinant. In the above, we have dropped the constant $\epsilon$ for simplicity.

2) For the case of $\mathcal{S} = \mathcal{U}$, we have

$$\Gamma_U := \Gamma_\mathcal{S}\big|_{|\mathcal{S}|=U} = d_U^{-1}\left(d_U(Ur)\right) = Ur$$

and (16) requires the overall code $\mathcal{C} = \mathcal{C}_1 \times \cdots \times \mathcal{C}_U$ to satisfy

$$\mathfrak{D}(\mathcal{U}) = \min_{D_u:\, u \in \mathcal{U}} \det\left(\Delta_C \Delta_C^\dagger\right) \;\dot\geq\; \mathrm{SNR}^0, \qquad (18)$$

where $\Delta_C = M(D_1, \cdots, D_U)$.

The above analysis shows that if $n_r = U n_t$ and if $r$ is in the antenna pooling regime, the criterion in Theorem 2 allows each user to use codes with vanishing determinant, but it expects the overall code $\mathcal{C}$ to satisfy the NVD criterion. A further investigation of this will show that the latter constraint actually leads to the total-NVD criteria for all subsets of users, and in particular that the individual codes must be NVD as well.

*Theorem 3:* For $H_{u,m} = H_u$, $n_r \in [Un_t, (U+1)n_t)$, $r_u = r \in \left[\frac{n_r}{U+1}, n_t\right]$, i.e. the system operates in the antenna pooling regime, the condition (5) of Theorem 2 for $\mathcal{S} \neq \mathcal{S}^*$ is the same as

$$\mathfrak{D}(\mathcal{S}) = \min_{D_u:\, u \in \mathcal{S}} \det\left(\Delta_C \Delta_C^\dagger\right) \;\dot\geq\; \mathrm{SNR}^0.$$

Thus (5) holds only for $0 \leq \gamma_\mathcal{S} \leq r|\mathcal{S}|$, and the resulting criterion is therefore called the *joint NVD criterion*.

*Proof:* First, let $\mathcal{S} = \mathcal{U}$ and let $\Delta_C = M(D_1, \cdots, D_U)$ be the matrix comprising of the difference matrices from all users. Fisher's inequality [9] on positive definite matrices shows

$$\det\left(\Delta_C \Delta_C^\dagger\right) \;\leq\; \prod_{u=1}^U \det\left(D_u D_u^\dagger\right).$$

Combining the above with (18) yields the condition

$$\mathrm{SNR}^0 \;\dot\leq\; \mathfrak{D}(\mathcal{U}) \leq \prod_{u=1}^U \min_{D_u} \det\left(D_u D_u^\dagger\right) = \prod_{u=1}^U \mathfrak{D}(\{u\}) \quad (19)$$

where the second inequality follows from the fact that the users do not cooperate. If every individual code has vanishing determinant, then the condition (18) cannot be satisfied. Thus there exists some user $v \in \mathcal{U}$ such that $\mathfrak{D}(\{v\}) = \min_{D_v} \det\left(D_v D_v^\dagger\right) \;\dot=\; \mathrm{SNR}^{p_v}$ for some $p_v > 0$. Note that user $v$ transmits at multiplexing gain $r$. Simply along the lines of the proof of approximately universal codes in [6] it can be shown that the single user DMT achieved by $\mathcal{C}_v$ is $d_1(r - p_v)$. This means user $v$ transmits at multiplexing gain $r$ and achieves diversity gain $d_1(r - p_v) > d_1(r)$, a contradiction to the point-to-point DMT. Therefore $p_v = 0$, and the above implies that every user $u$ achieves $\mathfrak{D}(\{u\}) \;\dot=\; \mathrm{SNR}^0$, i.e. the individual codes must be NVD, forced by the condition (18). The rest of the proof proceeds with induction on the size of $\mathcal{S}$. Assume we have shown for any $\mathcal{S} \subseteq \mathcal{U}$ with $|\mathcal{S}| = s$ and $\mathfrak{D}(\mathcal{S}) \;\dot\geq\; \mathrm{SNR}^0$. Let $\mathcal{S} = \mathcal{S}' \cup \{v\}$ and we will show the same NVD criterion holds for $\mathcal{S}'$. To establish this claim, again applying Fisher's inequality gives $\mathfrak{D}(\mathcal{S}) \leq \mathfrak{D}(\mathcal{S}') \cdot \mathfrak{D}(\{v\})$. As $\mathfrak{D}(\{v\}) \;\dot=\; \mathrm{SNR}^0$, we conclude that $\mathfrak{D}(\mathcal{S}') \;\dot\geq\; \mathrm{SNR}^0$. The claim on the range of $\gamma_\mathcal{S}$ follows obviously. ∎

The above theorem shows that when the symmetric MIMO-MAC system operates in the antenna pooling regime, condition (5) in Theorem 2 on the product of the minimum eigenvalues of the difference matrices $\Lambda_{m(\mathcal{S})}^{\rho|\mathcal{S}|n_t}(\mathrm{SNR})$ is equivalent to asking the minimum determinant $\mathfrak{D}(\mathcal{S})$ for all subsets of users to be nonvanishing. Furthermore, it shows the relaxation on $\gamma_\mathcal{S}$ is not possible. In the next section, we will further investigate the condition (16) and provide a general bound on $\mathfrak{D}(\mathcal{U})$ that can be applied to all values of $r$.

## III. GENERAL BOUNDS ON $\mathfrak{D}(\mathcal{U})$ FROM PIGEON HOLE PRINCIPLE

In this section we will restrict ourselves to the case of $n_r \geq U n_t$ receiving antennas. Assume that we are to design a MIMO-MAC system over a flat fading channel for $U$ users, each transmitting synchronously and independently with $n_t$ antennas at multiplexing gain $r$. We can describe each user's signals as $(n_t \times M)$ complex matrices with $M = U n_t$ required by Theorem 2. It is natural to assume that each user is maximally using all the degrees of freedom available to him/her. Therefore, the lattice of the individual code should be of full rank $n = 2 n_t M$ and the corresponding code $\mathcal{C}_u$ is given by

$$\mathcal{C}_u = \left\{ C_u = \sum_{k=1}^n a_{u,k} B_{u,k} : a_{u,k} \in \mathbb{Z}, -N \leq a_{u,k} \leq N \right\}$$

where the set $\{B_{u,k} : k = 1, \ldots, n\}$ is the basis of lattice $\mathcal{L}_u$ of user $u$. In other words, the parameter $a_{u,k}$ is the $2N$-PAM coordinate of the lattice $\mathcal{L}_u$ of user $u$. Equivalently, a QAM-oriented reader may then view $\mathcal{C}_u$ as a linear dispersion of $\frac{n}{2} = n_t M$ independently chosen $4N^2$-QAM symbols. As $|\mathcal{C}_u| = |2N|^{2n_t M}$, we shall set

$$N \;\dot=\; \mathrm{SNR}^{\frac{r}{2n_t}} \qquad (20)$$

such that user $u$ transmits at multiplexing gain $r$. The code $\mathcal{C}_u$ has average power $\mathbb{E}\|C_u\|^2 \;\dot=\; N^2 = \mathrm{SNR}^{\frac{r}{n_t}}$, hence a scaling constant $\kappa^2 = \mathrm{SNR}^{1 - \frac{r}{n_t}}$ is need such that $\kappa \mathcal{C}_u$ meets the power constraint $\mathbb{E}\|\kappa C_u\|^2 \;\dot=\; \mathrm{SNR}$. Moreover, it should be noted that the code $\mathcal{C}_u$ is the same as that discussed in the channel model (13). For any such code $\mathcal{C}_u$ and for any number of users, in this section we will aim to provide an upper bound on $\mathfrak{D}(\mathcal{U}) = \min_{D_u : u \in \mathcal{U}} \det\left(\Delta_C \Delta_C^\dagger\right)$ of condition (16).

To describe the idea we lead off with the simpler case $n_t = 1$, where user uses only single antenna, and is thus transmitting a vector $\underline{c}_u \in \mathcal{C}_u \subset \mathcal{L}_u \in \mathbb{C}^U$ since $M = U$. Set $\underline{d}_u = \underline{c}_u - \underline{c}'_u$ for some $\underline{c}_u \neq \underline{c}'_u \in \mathcal{C}_u$. Let us fix the difference signals $\underline{d}_u$ for all but one user, say, fix the vectors $\underline{d}_2, \cdots, \underline{d}_U$. We want to keep the coefficients $\{a_{u,k}, u = 2, \cdots, U; k =$

$1, \cdots, n\}$ of these as small as possible. If $\underline{d}_2, \cdots, \underline{d}_U$ are linearly dependent, then $\mathfrak{D}(\mathcal{U}) = 0$. We shall assume that these vectors form a linearly independent set. Therefore they span a complex vector space $W$ of dimension $U - 1$.

We shall be applying the pigeon hole principle in the quotient space $V = \mathbb{C}^U/W$. To make the calculations more specific we may identify $V$ with an orthogonal complement (with respect to the Euclidean inner product of $\mathbb{C}^U$ identified with $\mathbb{R}^{2U}$) of $W$ in $\mathbb{C}^U$. The mapping $f$ from $\mathbb{C}^U \to \mathbb{C}$ given by $f : \underline{c}_1 \mapsto \det M(\underline{c}_1, \underline{d}_2, \cdots, \underline{d}_U)$ is linear in $\underline{c}_1$ with constant coefficients of a bounded size (as we selected $\underline{d}_2, \cdots, \underline{d}_U$ with minimal coefficient $a_{u,k}$). Furthermore, $f(\underline{c}_1 - \underline{c}_1') = 0$ whenever $(\underline{c}_1 - \underline{c}_1') \in W$, so we can view $f$ as a linear function from $V$ to $\mathbb{C}$. Let $\pi : \mathbb{C}^U \to V$ be the natural projection that we may also think of as an orthogonal projection, i.e. a mapping that can only shrink a vector in length.

The assumption about the rank of the lattice $\mathcal{L}_1$ says that there are $\mathcal{O}(N^{2U})$ code vectors in $\mathcal{C}_1$. The coordinates of all of them in $\mathbb{C}^U$ are of the size $\mathcal{O}(N)$. As $\pi$ is a shrinking map, the coordinates of their images in $V$ are also of the size $\mathcal{O}(N)$, so they fall into a square shaped region $R \subset V$ with side length $\mathcal{O}(N)$, as a real vector space $V$ has dimension 2. Therefore we can partition the set $R$ into at most $|\mathcal{C}_1| - 1 = \mathcal{O}(N^{2U})$ smaller squares with side length $\mathcal{O}(N/N^U) = \mathcal{O}(1/N^{U-1})$. The pigeon hole principle then tells us that there exists a pair of distinct vectors $\underline{c}_1 \neq \underline{c}_1' \in \mathcal{C}_1$ such that $\pi(\underline{c}_1)$ and $\pi(\underline{c}_1')$ fall into the same small square. By linearity, the projection of their difference vector $\pi(\underline{d}_1) = \pi(\underline{c}_1) - \pi(\underline{c}_1')$ then has coordinates of size $\mathcal{O}(1/N^{U-1})$. Therefore also $f(\underline{d}_1) = \det M(\underline{d}_1, \underline{d}_2, \cdots, \underline{d}_U) = \det(\Delta_C)$ with $\Delta_C = M(\underline{d}_1, \underline{d}_2, \cdots, \underline{d}_U)$ has value of the size $\mathcal{O}(1/N^{U-1})$. We have proven the following result.

*Theorem 4:* (Pigeon hole bound, single antenna case) For any MIMO-MAC lattice code of $U$ users, each transmitting at multiplexing gain $r$ with $n_t = 1$, then there exists a constant $K > 0$ such that

$$\mathfrak{D}(\mathcal{U}) \leq \left|\frac{K}{N^{U-1}}\right|^2 \doteq \text{SNR}^{-(U-1)r}.$$

■

We turn our attention to the case of multiple transmit antennas. The application of the pigeon hole principle is very similar in spirit. We simply need to keep track of the dimesions of various vector spaces.

Let us, again, begin by fixing non-zero difference signal matrices $D_u = C_u - C_u'$ with $C_u \neq C_u' \in \mathcal{C}_u$ for users $u = 2, 3, \ldots, U$. We want to find a large subspace $W \subseteq \mathcal{M}_{n_t \times M}(\mathbb{C})$ with $M = Un_t$ such that $\det M(D, D_2, D_3, \ldots, D_U) = 0$ whenever $D \in W$. We assume $M(D_2, \cdots, D_U)$ is of full rank and the $(U-1)n_t$ rows of the blocks $D_2, \ldots, D_U$ are linearly independent, otherwise $\mathfrak{D}(\mathcal{U}) = 0$. Let $W'$ be their complex linear span, and let $W$ be the vector space consisting of complex $(n_t \times M)$ matrices with rows in $W'$. Obviously $\dim W = n_t \cdot \dim W' = (U-1)n_t^2$, so the quotient space $V = \mathcal{M}_{n_t \times Un_t}/W$ has real dimension $2n_t^2$. When we restrict the selection of $D_1 = C_1 - C_1'$ into $\mathcal{C}_1$ their projections in $V$ are confined to a hypercube $R$ of side length $\mathcal{O}(N)$. The size of the constellation $\mathcal{C}_1$ is $\mathcal{O}(N^{2Un_t^2})$ as the lattice $\mathcal{L}_1$ was assumed to be of a full rank $2Un_t^2$. Again we partition the $2n_t^2$-dimensional hypercube $R$ into $\mathcal{O}(N^{2Un_t^2})$ smaller cubes of side length $\mathcal{O}(N/N^U)$. As in the single antenna case we can then produce a non-zero difference vector $\underline{d}_1$ such that all the coordinates of its projection in $V$ are of the size $\mathcal{O}(1/N^{U-1})$. This time the determinant $f : C_1 \mapsto \det M(C_1, D_2, \cdots, D_U)$ is a polynomial function (with constant size coordinates) of degree $n_t$ of these coordinates of $C_1$, so by pigeon hole principle there must exist $C_1 \neq C_1' \in \mathcal{C}_1$ such that $f(D_1 = C_1 - C_1')$ has value $\left[\mathcal{O}(1/N^{U-1})\right]^{n_t}$. We have proven the following.

*Theorem 5:* (Pigeon hole bound, multi-antenna case). For any MIMO-MAC lattice code of $U$ users, each transmitting at multiplexing gain $r$ with $n_t =$ transmit antennas, there exists a constant $K > 0$ such that

$$\mathfrak{D}(\mathcal{U}) \leq \left|\frac{K}{N^{(U-1)n_t}}\right|^2 \doteq \text{SNR}^{-(U-1)r}.$$

■

Applying the above theorem to Theorem 3 we immediately see that there does not exist any MIMO-MAC codes satisfying condition (5) when the MIMO-MAC system operates in the antenna pooling regime. For the single-user performance regime, the dominant outage set $\mathcal{S}^* = \{1\}$ and condition (5) requires $\mathfrak{D}(\{1\}) \dot{\geq} \text{SNR}^0$ for the individual code and $\mathfrak{D}(\mathcal{U}) \dot{\geq} \text{SNR}^{-(\Gamma_\mathcal{U} - Ur - \epsilon)}$ for the overall code. However, Theorem 5 shows that this is possible only if $\Gamma_\mathcal{U} \geq (2U-1)r$, i.e. we need $d_\mathcal{U}((2U-1)r) \geq d_1(r)$.

*Corollary 6:* With $n_r \geq Un_t$, assume all users transmit at multiplexing gain $r$. Then

1) MIMO-MAC lattice codes satisfying (5) might exist only when $r \ll \frac{n_t}{2}$, and
2) there do not exist MIMO-MAC lattice codes satisfying (5) if $r \in [\frac{n_t}{2}, n_t]$.

■

## REFERENCES

[1] P. Coronel, M. Gärtner, and H. Bölcskei, "Diversity multiplexing tradeoff in selective fading multiple-access MIMO channels," http://arxiv.org/abs/0905.1386v2.
[2] ——, "Diversity multiplexing tradeoff in selective fading multiple-access MIMO channels," in *Proc. 2008 IEEE Int. Symp. Inform. Theory*, Toronto, ON, Jul. 2008.
[3] L. Zheng and D. Tse, "Diversity and multiplexing: a fundamental tradeoff in multiple antenna channels," *IEEE Trans. Inf. Theory*, vol. 49, no. 5, pp. 1073–1096, May 2003.
[4] H. F. Lu, "Explicit constructions of multi-block space-time codes that achieve the diversity-multiplexing tradeoff," *IEEE Trans. Inf. Theory*, vol. 54, no. 8, pp. 3790–3796, Aug. 2008.
[5] D. N. C. Tse, P. Viswanath, and L. Zheng, "Diversity-multiplexing tradeoff in multiple-access channels," *IEEE Trans. Inf. Theory*, vol. 50, no. 9, pp. 1859–1874, Sep. 2004.
[6] P. Elia, K. R. Kumar, S. A. Pawar, P. V. Kumar, and H.-F. Lu, "Explicit construction of space-time block codes achieving the diversity-multiplexing gain tradeoff," *IEEE Trans. Inf. Theory*, vol. 52, no. 9, pp. 3869–3884, Sep. 2006.
[7] J.-C. Belfiore and G. Rekaya, "Quaternionic lattices for space-time coding," in *Proc. IEEE Information Theory Workshop*, Paris, 31 March - 4 April 2003.
[8] S. Tavildar and P. Viswanath, "Approximately universal codes over slow fading channels," *IEEE Trans. Inf. Theory*, vol. 52, no. 7, pp. 3233–3258, Jul. 2006.
[9] R. A. Horn and C. R. Johnson, *Matrix Analysis*. Cambridge, UK: Cambridge University Press, 1985.